\documentclass[floatfix,prl,aps,twocolumn,showpacs,preprintnumbers,amsmath,amssymb]{revtex4}

\usepackage{graphicx}
\usepackage{dcolumn}
\usepackage{bm}

\begin{document}
\title{Surface Effects in Magnetic Microtraps}
\author{J. Fort\'{a}gh}
\email{fortagh@pit.physik.uni-tuebingen.de}
\author{H. Ott}
\author{S. Kraft}
\author{A. G\"unther}
\author{C. Zimmermann}

\affiliation{Physikalisches Institut der Universit\"at T\"ubingen\\
Auf der Morgenstelle 14, 72076 T\"ubingen, Germany}

\begin{abstract}
We have investigated Bose-Einstein condensates and ultra cold
atoms in the vicinity of a surface of a magnetic microtrap. The atoms are prepared along copper conductors at distances to the surface between 300 $\mu$m
and 20 $\mu$m. In this range, the lifetime decreases from 20 s to
0.7 s showing a linear dependence on the distance to the surface.
The atoms manifest a weak thermal coupling to the surface, with
measured heating rates remaining below 500 nK/s. In addition, we
observe a periodic fragmentation of the condensate and thermal clouds when the surface is approached.

\end{abstract}

\maketitle

Micropotentials have proven to be a powerful tool to manipulate
and structure the shape of a Bose-Einstein condensate
\cite{Inguscio1999a} on a length scale shorter than the coherence
length of the condensate. Besides the manipulation with light
\cite{Andrews1997b,Anderson1998a,Cataliotti2001a,Greiner2001a,Greiner2002a},
current carrying microstructures \cite{Weinstein1995a} are
particularly interesting since they can be tailored in an
arbitrary way, providing a variety of potential geometries. In
previous experiments with magnetic microtraps the work was mainly
focused on the demonstration of different trapping geometries,
loading schemes and guiding principles
\cite{Vuletic1998a,Fortagh1998a,Reichel1999a,Muller1999a,Dekker2000a,Key2000a,Cassettari2000a}.
The recent realization of Bose-Einstein condensates in magnetic
microtraps \cite{Ott2001a,Hansel2001b} however provides new
possibilities to control coherent matter on the micrometer scale.
Coherent beam splitters, on-chip interferometers or quantum dots
may become feasible. In current experimental setups, the
dimensions of the conductors vary from 1 $\mu$m to 100 $\mu$m and
the distance between the trap minimum and the surface is typically
of the same size. At such small distances the atoms are affected
by the nearby surface. For experiments with coherent matter waves
or even single atoms in microfabricated traps an understanding of
the mutual influences of the atoms and the surface is highly
desirable.

In this letter, we describe three effects on ultra cold atoms
which appear in the vicinity of the surface of a magnetic
microtrap. We observe a decrease of the lifetime of the atomic
cloud which scales roughly linearly with the distance to the
surface. At 20 $\mu$m, the lifetime is reduced to less than 1 s,
which has to be compared to the ``far distance'' value of 100 s.
Simultaneously, an increased heating rate is observed which, however, does
not exceed 500 nK/s. Furthermore, a periodic fragmentation of both,
the thermal cloud and the condensate occurs when the surface is
approached at distances of about 250 $\mu$m. This gives strong
evidence for additional potentials arising from the nearby surface.

In our experiment, the microtrap is generated by a microstructure
which consists of seven parallel copper conductors with widths of
3 $\mu$m, 11 $\mu$m and 30 $\mu$m, a height of 2.5 $\mu$m and a
length of 25 mm \cite{For2}. The conductors are electroplated on
an Al$_2$O$_3$ ceramic substrate. An additional copper wire with a circular 
diameter of 90 $\mu$m is mounted parallel to the microstructure,
allowing for reference measurements. The free surface of the wire is in the plane of the fabricated conductors. The radial potential is built
from the field of one of the eight conductors with current $I$ and
a homogeneous bias field $B_{\mathrm{bias}}$ perpendicular to the
conductor. The axial confinement is accomplished by a superimposed
Ioffe type trap \cite{Fortagh2000a}, that also provides a
non-vanishing field $B_0$ in the centre of the trap. The axial
oscillation frequency has an upper limit of $2\pi\times14$ Hz and
can be tuned without affecting the radial confinement by changing
the strength of the Ioffe type trap. In all experiments, a
pre-cooled cloud of $^{87}$Rb atoms in the $|F=2,m_F=2>$ hyperfine
ground state is transferred into the microtrap. The loading
procedure, the adiabatic transfer, and the experimental cycle are
described in detail elsewhere \cite{Ott2001a}. The trap geometry
is characterized by its axial and radial oscillation frequency
$\omega_a$ and $\omega_r$, and by its distance $d$ to the surface. 
Assuming a linear conductor carrying current and a homogeneous bias field perpendicular to the conductor, $\omega_r$ and $d$ are given by
\begin{equation*}
\omega_r=2\pi\times\frac{1}{\mu_0}\sqrt{\frac{\mu_Bg_Fm_F}{mB_0}}\times\frac{B_{\mathrm{bias}}^2}{I},
\end{equation*}
\begin{equation*}
d=\frac{\mu_0}{2\pi}\times\frac{I}{B_{\mathrm{bias}}},
\end{equation*}
where m denotes the mass of the atom. For a given distance $d$ and
a radial oscillation frequency $\omega_r$ the current $I$ in the
conductor and the required bias field $B_{\mathrm{bias}}$ are
fully determined. Thus, the radial steepness of the waveguide and
its distance to the surface can be tuned independently.

\begin{figure} \centering
\includegraphics[width=6.8cm]{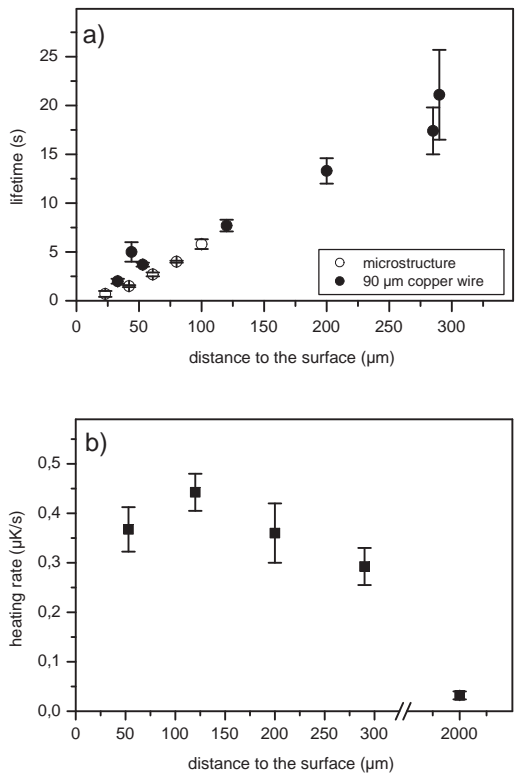}
\caption{\label{lifetime} Lifetime and heating rate of thermal
atoms in the vicinity of the surface ($T=1.5$ $\mu$K). a) Lifetime
near the surface of the microstructure and the 90 $\mu$m wire. The
lifetime for a distance of 2 mm is $100 s$. b) Heating rate at the
90 $\mu$m copper wire. Measurements at the microstructure yield to
similar results. The last data point on the right is taken at a
distance of 2 mm.}
\end{figure}

To investigate lifetime and heating rate we
prepare cold thermal ensembles at different distances to the trap surface. The temperature is adjusted to 2 $\mu$K, well
above the critical temperature for Bose-Einstein condensation. The
trap is kept constant for a variable storage time and the number
of atoms as well as the temperature are determined from time of
flight measurements of the released atomic cloud.
Fig.~\ref{lifetime}a shows the measured trap lifetimes at the
microstructure and at the 90 $\mu$m copper wire. The lifetime
reveals a linear dependence on the distance to the conductor
surface over a wide range. For small distances the data differ
from a pure linear behaviour. Both types of conductor (massive
copper wire and thin, electroplated conductor) show a similar
influence such that close to the conductor the lifetime is reduced
by two orders of magnitude. For a distance of 20 $\mu$m the
measured lifetime is as short as 700 ms, whereas for a distance of
2 mm a reference value of 100 s has been determined. 
To bring the atoms close to the surface the current in the
conductor are reduced. In this case, the dissipated heat in the conductor due to
resistive heating scales as $P\propto d^4$. Thus, at smaller distances the
temperature of the conductor is lower. In our experiment, the
conductors are mounted on a cooled copper heat sink which provides an
almost constant temperature. To avoid influences due to
long-time heating of the equipment the data points for different distances
to the surface were measured in random order. This allows us to exclude
outgassing effects. Evaporation of cold atoms that hit the surface
can also be excluded: at small distances the potential barrier at the
conductors surface is approximately 6 G corresponding to a temperature of
more than 400 $\mu$K.
In
\cite{Henkel1999a} surface induced spin flip transitions are
analyzed as lifetime reducing mechanism. Spin flips can be induced
by the oscillating magnetic field of thermally excited currents 
in the metal or by technical noise at frequencies around 1 MHz.
Because the radiation field of a dipolar antenna decreases inversely 
proportional with the distance the measured data are in
qualitative agreement with the theory. 

The experiments were performed on thermal atomic clouds at densities
below $1\times10^{14}\,\mathrm{cm}^{-3}$ and no indication of inelastic
processes like three body recombination \cite{Burt1997a} was
observed. The generation of Bose-Einstein condensates results in
higher densities and inelastic processes become important. In
highly elongated traps with $\omega_a=2\pi\times14$ Hz and
$\omega_r>2\pi\times1000$ Hz as used for 
the previous experiments, we reach condensation at densities
above $1\times10^{15}\,\mathrm{cm}^{-3}$ and the lifetime of the condensate
is limited to a few 10 ms. Longer lifetimes can be achieved if the
density is reduced by adiabatic relaxation of the trapping
potential. For $\omega_a=2\pi\times8$ Hz, $\omega_r=2\pi\times100$
Hz and d=300 $\mu$m we measure a condensate lifetime of 2.3 s.

The heating rate has been determined by measuring the temperature
of the cloud after different storage times. Fig.~\ref{lifetime}b
shows the heating rate for different distances to the surface. In
all studied scenarios, the heating rate was less than 500 nK/s.
For comparison, at a distance of 2 mm a reference value of
$32\pm8$ nK/s has been determined, indicating an extremely low
technical noise in our apparatus. Thus, the increased heating rate
can be addressed to specific effects close to the conductors surface.
Heating can be induced due to fluctuations of the magnetic field at the trap frequencies, at their half or harmonics, however we found no indication for such kind of noise in our experiment. 
Although the lifetime of atomic clouds close to the surface 
is significantly reduced, this puts only moderate restrictions
on applications of magnetic microtraps. 
For a condensate the heating leads to additional losses. However, it 
does not necessarily  reduce the coherence properties.

More important, we have observed an unexpected periodic fragmentation of the atomic cloud
along the conductors at distances of less than approximately 250 $\mu$m. 
\begin{figure}
\centering
\includegraphics[width=6.8cm]{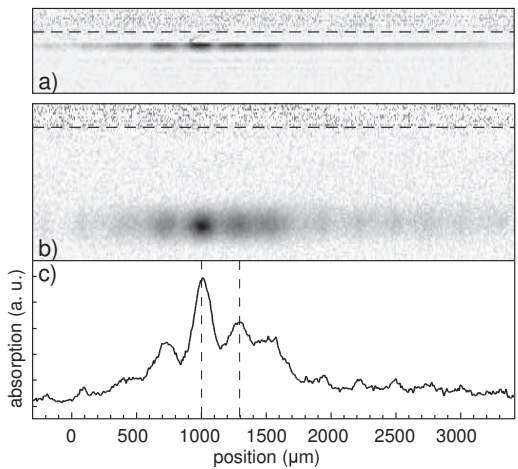}
\caption{\label{surface}Fragmentation of a thermal cloud in the
vicinity of the surface ($N=560,000$, T=1 $\mu$K). The axial
confinement was ramped down to zero within 400 ms and the cloud
was allowed to freely expand for 100 ms. The absorption images
were taken a) in the waveguide and b) after 10 ms time of flight.
c) The integrated scan shows a periodicity of $300$ $\mu$m in the
density distribution of the cloud along the waveguide. The radial
oscillation frequency was $\omega_r=2\pi\times1000$ Hz and the
waveguide potential was located at a distance $d=150$ $\mu$m to
the surface. The dashed line indicates the surface of the
microtrap and the orientation of the conductors.}
\end{figure}
To study this fragmentation we have prepared a cloud of ultra cold
atoms at different distances to the surface. Subsequently, the
axial confinement of the trap was turned off within 400 ms and a
homogeneous bias field was applied along the axial direction in
order to retain the parabolic character of the radial confinement.
This configuration allows for a free propagation of the atoms in
the waveguide. An undisturbed propagation is observed for
distances larger than 300 $\mu$m. Below 250 $\mu$m, the guiding
potential obviously exhibits a significant waviness. Fig.~\ref{surface} shows
the distribution of a thermal cloud after 100 ms of expansion in
the waveguide. The modulation of the atomic density distribution
indicates the presence of an additional periodic trapping potential generated by the microstructure. The spacing
between the potential minima is 300 $\mu$m and the depth of the
potential is on the order of $k_B\times1\,\mu\mathrm{K}$ which can be
estimated by the temperature of the atomic cloud. Changing the
distance to the surface has no influence on the position and the
period of the potential minima. Releasing a Bose-Einstein
condensate in the waveguide allows for the detection of finer
substructures. Fig.~\ref{twobec}a shows two condensates with a
separation of 110 $\mu$m. The image was taken 300 ms after the
axial relaxation was completed and the waveguide potential was
located 100 $\mu$m below the microstructure. The axial
confinement of the microtrap was completely turned off and the
condensates were still trapped in two potential minima at the
surface. An even finer structure can be observed at a distance of
50 $\mu$m. Fig.~\ref{twobec}b shows the formation of individual
condensates with a separation of 50 $\mu$m. In this experiment,
the image was taken directly after the last cooling stage in the
trap without ramping down the axial confinement.

The appearance of the surface potentials can be observed at each
of the seven conductors of the microstructure. Dependencies of
the period and position of the potential structure have to be investigated in future experiments.
The experiments were repeated at the 90 $\mu$m
copper wire. Here, we observe a periodic modulation with a spacing
of 220 $\mu$m including a substructure with 110 $\mu$m. 

\begin{figure}
\centering
\includegraphics[width=6.8cm]{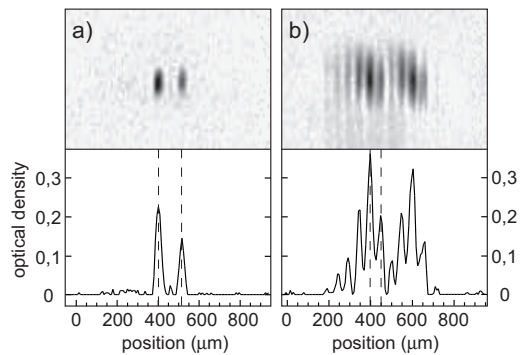}
\caption{\label{twobec} Fragmentation of a Bose-Einstein
condensate close to the surface. The absorption images were taken
after 15 ms time of flight. a) Two condensates 300 ms after
relaxation of the axial confinement. The spacing of the
condensates is $110$ $\mu$m ($d=100$ $\mu$m). The condensate at
the left contains 12,000 atoms, at the right 7,000 atoms. b)
Multiple condensates with a periodic spacing of $50$ $\mu$m
($d=50$ $\mu$m). The condensates were released from the trap
without relaxation of the axial confinement. The envelope shows a
$200$ $\mu$m structure which is due to the superposition of the
surface potential with a period of $300$ $\mu$m
(Fig.~\ref{surface}) and the axial confinement
($\omega_a=2\pi\times14$ Hz) of the microtrap. The total number of
atoms is 165,000, the tallest part contains 33,000 atoms.}
\end{figure}

\begin{figure}
\centering
\includegraphics[width=6.6cm]{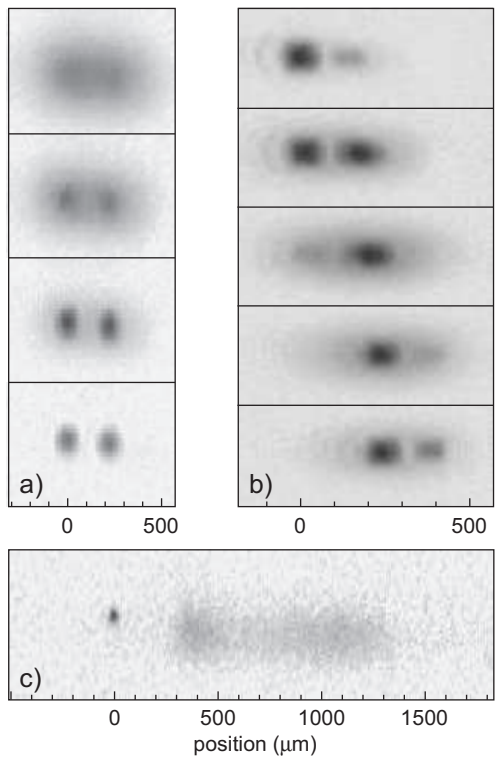}
\caption{\label{pinning} Pinning of the condensates. a)
Simultaneous formation of two condensates at a distance $d=150$
$\mu$m to the surface. The superposition of the surface potential
and the axial confinement of the trap results in a spacing of
$220$ $\mu$m between the two potential minima. The images were
taken at different temperatures below $T_c$ after 15 ms time of
flight. The critical temperature is reached with 500,000 atoms
(first image), the condensates contain 65,000 atoms each. b) By
shifting the centre of the trap along the waveguide, the
condensates populate stationary potential minima (10 ms time of
flight). c) Separation of a condensate (5,000 atoms) and the
thermal cloud (90,000 atoms) after 100 ms of free expansion in the
waveguide (15 ms time of flight).}
\end{figure}

To exclude interatomic interaction as a possible reason for the
separation, we have generated two neighboring condensates as shown
in Fig.~\ref{pinning}a. One of them was subsequently removed with a
focused laser beam. In accordance with our assumption of periodic
surface potentials the second condensate remains at the same
position, whereas a fraction of the thermal cloud starts to
oscillate in the trap.

Magnetic modulations caused by a transverse current variation
decay on a length scale which is on the order of the transverse
dimension of the conductor. Thus, modulations over larger
distances can only be addressed to longitudinal variations. They
cannot arise from the flow of the electric current, which has to be
constant along the conductor. To exclude any material impurities
like ferromagnetic domains, we have made a chemical analysis
(energy dispersive X-ray analysis, EDX) of an identical
microstructure, which revealed no measurable impurities
\cite{For2}. Trivial causes for modulated electrostatic or
gravitational potentials can be excluded. One may speculate that the observed potential modulations could be attributed to 
longitudinal or transversal spin arrangements of moving electrons 
in copper \cite{Hirsch1999a}.
Meanwhile, the observations discussed above could be reproduced by two other groups \cite{lean,jones} working on microtraps.
The common feature of the experiments is the use of 
copper conductors as current-carrying elements for generating 
the micro trap, with differences due to the various 
conductor profiles and currents.

The appearance of additional potentials at the surface has several
consequences for the application of magnetic microtraps for
Bose-Einstein condensates. A continuous shift of the condensate
along the conductor or an undisturbed propagation in a waveguide
is only possible at a sufficiently large distance to the surface.
Fig.~\ref{pinning}b shows a sequence of absorption images where the
trapping potential was shifted in the axial direction. As a result
the condensates are pinned on the periodic potential and populate
subsequent potential minima. Another example for the pinning
effect is illustrated in Fig.~\ref{pinning}c. The atomic ensemble
is shown after 100 ms of free axial expansion in the waveguide.
The condensate remains pinned at its original position while the
thermal cloud starts to propagate in the waveguide resulting in a
separation of the two components.

In conclusion, we have investigated the lifetime and the heating
rate of ultra cold gases in the vicinity of a surface for a set of
parameters that corresponds to currently used microtrap setups. We
find evidence for influences of the surface resulting in a
reduction of the lifetime and an increased heating rate. The
appearance of a periodic potential near the surface is a
qualitatively new phenomenon which has to be investigated in
future experiments.

The authors would like to thank G. Mih\'{a}ly for helpful
discussions. This work was supported in part by the Deutsche
Forschungsgemeinschaft.

\end{document}